\documentclass[twocolumn,prb]{revtex4-2}
\usepackage{epsfig,amssymb,amsmath}
\usepackage{graphicx}
\usepackage{color}
\usepackage[english]{babel}

\begin{document}

\title{Nonlinear features of the superconductor--ferromagnet--superconductor  $\varphi_0$ Josephson junction in ferromagnetic resonance region}

\author{Aliasghar Janalizadeh$^{1}$}
\author{Ilhom R. Rahmonov$^{2,3,4}$}
\author{Sara A. Abdelmoneim~$^{5}$}
\author{Yury M. Shukrinov$^{2,3,4}$}
\author{Mohammad R. Kolahchi$^{1}$}

\address{$^{1}$ Department of Physics, Institute for Advanced Studies in Basic Sciences (IASBS), P.O. Box 45137-66731, Zanjan, Iran\\
$^{2}$ BLTP, JINR, Dubna, Moscow Region, 141980, Russia\\
$^{3}$ Dubna State University, Dubna,  141980, Russia\\
$^{4}$ Moscow Institute of Physics and Technology, Dolgoprudny, 141700, Moscow Region, Russia\\
$^{5}$ Physics department, Menofiya University,  Faculty of Science, 32511, Shebin Elkom,Egypt
}

\date{\today }

\begin{abstract}
We demonstrate the manifestations of the nonlinear features in magnetic dynamics and IV-characteristics  of the $\varphi_0$ Josephson junction in the ferromagnetic resonance region. We show that at small values of system parameters, namely,  damping, spin-orbit interaction, and Josephson to magnetic energy ratio, the magnetic dynamics is reduced to the dynamics of the scalar Duffing oscillator, driven by the Josephson oscillations. The role of increasing superconducting current in the resonance region is clarified. Shifting of  the ferromagnetic resonant frequency and the reversal of its damping dependence due to nonlinearity are demonstrated by the full Landau-Lifshitz-Gilbert-Josephson system of equations, and in its different approximations. Finally, we demonstrate the negative differential resistance in the IV--characteristics, and its correlation with the foldover effect.
\end{abstract}


\maketitle

\section{I. Introduction}
The coupling of superconducting phase difference with magnetic moment of ferromagnet in the $\varphi_0$ junction leads to a number of unique features important for superconducting spintronics, and modern information technology \cite{buzdin-prl08,linder15,bobkova18,bobkova20,szombati16}. It allows to control the magnetization precession by superconducting current and affects the current--voltage (IV) characteristics by magnetic dynamics in the ferromagnet, in particular, to create a DC component in the superconducting current \cite{konschelle-prl09,shukrinov-prb19,shukrinov-ufn21}.  A remarkable manifestation of such coupling is the possibility to stimulate a magnetization reversal in the ferromagnetic layer by applying  current pulse through the $\varphi_0$-junction \cite{bobkova18,linder11,hoffman12,eschrig15,shukrinov-apl17,mazanik-pra20}.

There are two features of our Josephson junction that come into play in our study. One is the broken inversion symmetry in the weak link of the Josephson junction, when the link is magnetic, which introduces an extra phase in the current—-phase relation, preventing it from being antisymmetric. Such Josephson junctions are named $\varphi_0$ junctions \cite{buzdin-prl08}, and examples exist such as MnSi and FeGe. Second is the nonlinear property of the system that makes for an anomalous resonance behavior \cite{shukrinov-prb21}.

We couple such a Josephson junction to the model that describes the magnetodynamics in thin films or heterostructure, to form the Landau-Lifshitz-Gilbert-Josephson model (LLGJ)\cite{shen2020current,nik20,shukrinov-prb21}. It is shown that for a particular set of parameters, the coupled equations reduce to the dynamics of a Duffing oscillator \cite{shukrinov-prb21}. The cubic nonlinearity in this oscillator has applications in describing several effects in other models too \cite{moon15}. One being the resonance effects in the antiferromagnetic bimeron in response to an alternating current, which has applications in the detection of weak signals \cite{shen2020current,wan99,alm07}.

The Gilbert damping term is added phenomenologically to the Landau—-Lifshitz model, to reproduce the damping of the precessing magnetic moment. Gilbert damping is important in modeling other resonance features too, as its temperature dependence affects them \cite{zhao16,yao2018}, and in return in the superconducting correlations that affect it \cite{silaev2020}. The magnetization precession in the ultra thin $Co_{20}Fe_{60}B_{20}$ layer stimulated by microwave voltage under a large angle, needs modeling by Duffing oscillator too. This gets help from the so called foldover features, again due to nonlinearity \cite{nik20,nay79,chen09}.

The consequences of the nonlinear nature of the coupled set of LLGJ system of equations in the weak coupling regime was demonstrated recently in Ref. \cite{shukrinov-prb21}. We showed in this regime, where the Josephson energy is small compared to the magnetic energy, the $\varphi_0$ Josephson junction is equivalently described by a scalar nonlinear Duffing equation. An anomalous dependence of the ferromagnetic resonant frequency (FMR) with the increase of the Gilbert damping was found. We showed that the damped precession of the magnetic moment is dynamically driven by the Josephson supercurrent, and the resonance behavior is given by the Duffing spring. The obtained results were based on the numerical simulations. The role of dc superconducting current, and the state with negative differential resistance (NDR) in IV-characteristic were not clarified. Also, the effects of the Josephson to magnetic energy ratio and the spin-orbit coupling (SOC) were not investigated at that time.

In the present paper, we study the nonlinear aspects of the  magnetic dynamics and IV-characteristics  of the $\varphi_0$ Josephson junction in the ferromagnetic resonance region. We compare description of the anomalous damping dependence (ADD) exhibited by full LLGJ system of equations with approximated equations and demonstrate the Duffing oscillator features in the small parameter regime. Effects of the Josephson to magnetic energy ratio, and the spin-orbit coupling on the ADD, referred to earlier as the $\alpha$-effect \cite{shukrinov-prb21} are demonstrated. By deriving the formula which couples the dc superconducting current and maximal amplitude of magnetization we discuss the correlation of superconducting current and the negative differential resistance in the resonance region.  Finally, we discuss the experimentally important features by emphasizing the details of the magnetization dynamics and the IV-characteristics of the $\varphi_0$ junction.

We have shown that in the limit of small system parameters; that is, the Josephson to magnetic energy ratio
$G$, the damping $\alpha$, and the spin-orbit coupling $r$, the dynamics is given by the Duffing
spring \cite{shukrinov-prb21}. We focus on the shift in resonance and the effects of nonlinear interactions. We give semi-analytic models to explain our results in various limits.

The paper is organized as follows. In Section II we outline the theoretical model and discuss the methods of calculations. The ferromagnetic resonance and effects of system parameters on the anomalous damping dependence are considered in Subsection A of Section III.
In Subsection B we present analytical description of the dynamics and IV-characteristics of the $\varphi_{0}$ junction at small system parameters. Manifestation of the negative differential resistance in IV-characteristics through the foldover effect is discussed. We compare the description of the anomalous damping dependence by full LLGJ system of equation with approximated equation, and show how the Duffing oscillator captures the nonlinearities in the small parameter regime in Subsection C. We present results on the critical damping and derive the formula which couples the dc superconducting current and maximal amplitude of magnetization in the ferromagnetic layer.  Finally, in Section IV we concludes the paper.

\section{II. Models and Method}

\begin{figure}
	\includegraphics[height=35mm]{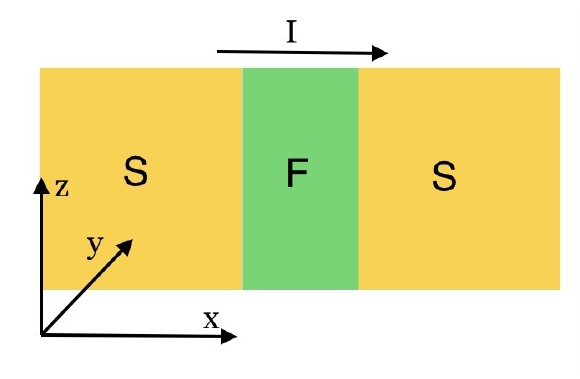}
	\caption{Schematic view of SFS $\varphi_0$ Josephson junction. The external current applied along x direction, ferromagnetic easy axis is along z direction.}
	\label{1}
\end{figure}

The following section is closely related to our work in~\cite{mazanik-pra20}. The $\varphi_0$ junction~\cite{guarcello20,shukrinov-apl17,konschelle-prl09} that we study is shown in Fig.\ref{1}. The current-phase relation  in $varphi_{0}$ junction has the form  $I_s = I_c \sin{(\varphi - \varphi_0)}$, where $\varphi_0 = r M_y/M_0$, $M_y$ denotes the component of magnetic moment in $\hat{y}$ direction, $M_0$ is the modulus of the magnetization. The physics of $\varphi_{0}$ Josephson juncton is determined by system of equations which consists of Landau-Lifshits-Gilbert (LLG), resistively capacitively shunted junction (RCSJ) model expression with current-phase relation ($I_s$) described above, and Josephson relation between phase difference and voltage.

The dynamics of the magnetic moment $\mathbf{M}$ is described by the LLG equation \cite{landau}

\begin{equation}\label{eq:LLG}
	\frac{d \mathbf{M}}{dt} = -\gamma \mathbf{M}\times\mathbf{H}_{eff} + \frac{\alpha}{M_0} \left(\mathbf{M}\times \frac{d \mathbf{M}}{dt}\right),
\end{equation}
where $\mathbf{M}$ is the magnetization vector, $\gamma$ is the gyromagnetic relation, $\mathbf{H}_{eff}$ is the effective magnetic field, $\alpha$ is Gilbert damping parameter, $M_0=|\mathbf{M}|$.

In order to find the expression for the effective magnetic field we have used the model developed in Ref.\cite{konschelle-prl09}, where it is assumed that the gradient of the spin-orbit potential is along the easy axis of magnetization taken to be along ${\hat z}$. In this case the total energy of the system can be written as
\begin{equation}
	E_{\text{tot}}=-\frac{\Phi_{0}}{2\pi}\varphi I+E_{s}\left(  \varphi,\varphi_{0}\right)  +E_{M}\left(  \varphi_{0}\right)  , \label{EQ_energy}%
\end{equation}
where $\varphi$ is the phase difference  between the superconductors across the junction, $I$ is the external current,
$E_{s}\left( \varphi,\varphi_{0}\right)=E_{J}\left[  1-\cos\left( \varphi-\varphi_{0}\right)  \right]$, and $\displaystyle
E_{J}=\Phi_{0}I_{c}/2\pi$ is the Josephson energy. Here $\Phi_{0}$ is the flux quantum, $I_{c}$ is the critical current, $r=l\upsilon_{so}/\upsilon_{F}$	$l=4 h L/\hbar \upsilon_{F}$, $L$ is the length of $F$ layer, $h$ is the exchange field of the $F$ layer, $E_{M}=-K\mathcal{V}M^2_z/(2 M_0^2)$, the parameter $\upsilon_{so}/\upsilon_{F}$ characterizes a relative strength of spin-orbit interaction, $K$ is the anisotropic constant, and $\mathcal{V}$ is the volume of the ferromagnetic ($F$) layer.

The effective field for LLG equation is determined by
\begin{eqnarray} \label{eq:Heff}
	{\bf H_{eff}}&=&-\frac{1}{\mathcal V}\frac{\partial E_{tot}}{\partial
		{\bf M}}\nonumber\\
	&=&\frac{\Omega_{F}}{\gamma}\bigg[G r \sin\bigg(\varphi - r
	\frac{M_{y}}{M_{0}} \bigg) {\bf\widehat{y}} +
	\frac{M_{z}}{M_{0}}{\bf\widehat{z}}\bigg]
	\label{effective_field}
\end{eqnarray}
where $\Omega_{F}=\gamma K/M_{0}$ is frequency of ferromagnetic resonance and $G= E_{J}/(K \mathcal{V})$ determines the ratio of Josephson energy to magnetic one.

In order to describe the full dynamics $\varphi_{0}$ junction the LLG equations should be supplemented by the equation for phase difference $\varphi$, i.e. equation of RCSJ model for bias current and Josephson relation for voltage. According to the extended RCSJ model, which takes into account derivative of $\varphi_{0}$ phase shift, the current flowing through the system in underdamped case is determined by
\begin{eqnarray}\label{eq:RSJM}
	I =  \frac{\hbar C}{2e} \frac{d^{2} \varphi}{ d t^{2}}+\frac{\hbar}{2eR}\left[ \frac{d \varphi}{ d t} - \frac{r}{M_{0}} \frac{d M_y}{d t}\right]\\
	+ I_{c}\sin{\left(\varphi - \frac{r}{M_{0}} M_y\right)}\nonumber.
\end{eqnarray}
where $I$ is the bias current, $C$ and $R$ are the capacitance and resistance of Josephson junction respectively. The Josephson relation for voltage is given by :
\begin{equation}
	\label{eq:JR}
	\frac{\hbar }{2e} \frac{d \varphi}{ d t}= V.
\end{equation}

We note that in the framework of RCSJ--model the displacement current is proportional to the first derivative of voltage (or second derivative of phase difference). From the other hand, the magnetization dynamics plays role of the external force and first order derivative of $\varphi_{0}$ is a source of external current for JJ. This  was demonstrated in  Ref.\cite{rabinovich2019, guarcello20} where the authors  included the first derivative of $\varphi_{0}$ as the source of the electromotive force. Voltage is determined by the phase difference, and does not depend on $\varphi_{0}$. From this point of view, in the framework of RCSJ model the external current source cannot modify the expression for displacement current. That's why we do not include the second derivative of $varphi_{0}$ in our model.

Using (\ref{eq:LLG}), (\ref{effective_field}), (\ref{eq:RSJM}) and (\ref{eq:JR}) we can write the  system of equations, in normalised variables, which describes the dynamics of $\varphi_{0}$ junction
\begin{equation}
	\label{syseq}
	\begin{array}{llll}
		\displaystyle \dot{m}_{x}=\frac{\omega_{F}}{1+\alpha^{2}}\{-m_{y}m_{z}+Grm_{z}\sin(\varphi -rm_{y})\\
		-\alpha[m_{x}m_{z}^{2}+Grm_{x}m_{y}\sin(\varphi -rm_{y})]\},
		\vspace{0.2 cm}\\
		\displaystyle \dot{m}_{y}=\frac{\omega_{F}}{1+\alpha^{2}}\{m_{x}m_{z}\\
		-\alpha[m_{y}m_{z}^{2}-Gr(m_{z}^{2}+m_{x}^{2})\sin(\varphi -rm_{y})]\},
		\vspace{0.2 cm}\\
		\displaystyle \dot{m}_{z}=\frac{\omega_{F}}{1+\alpha^{2}}\{-Grm_{x}\sin(\varphi -rm_{y})\\
		-\alpha[Grm_{y}m_{z}\sin(\varphi -rm_{y})-m_{z}(m_{x}^{2}+m_{y}^{2})]\},
		\vspace{0.2 cm}\\
		\displaystyle \dot{V}=\frac{1}{\beta_{c}}[I-V+r\dot{m}_{y}-\sin(\varphi-r m_{y})],\\
		\vspace{0.05 cm}\\
		\displaystyle \dot{\varphi}=V
	\end{array}
\end{equation}
where $m_{x,y,z} = M_{x,y,z}/M_0$ and satisfy the constraint $\sum_{i=x,y,z} m_{i}^2(t)=1$, $\beta_{c} = 2eI_{c}C R^{2}/\hbar$ is McCumber parameter. In order to use the same time scale in the LLG and RCSJ equations in this system of equations we have normalized time to the $\omega^{-1}_c$, where $\omega_{c}=\frac{2 e I_c R}{\hbar}$, and $\omega_{F}=\Omega_{F}/\omega_{c}$ is the normalized frequency of ferromagnetic resonance $\Omega_{F}=\gamma K/M_{0}$. Bias current is normalized to the critical current $I_{c}$ and voltage $V$ -- to the $V_{c}=I_{c}R$. The system of equations (\ref{syseq}), is solved numerically using the fourth-order Runge-Kutta method(see Ref.\cite{shukrinov-prb21}).

\section{III. Results and Discussion}

\subsection{A. Effect of system parameters on the anomalous damping dependence}

ADD of the FMR frequency with increasing $\alpha$ was discussed in Ref. \cite{shukrinov-prb21}. It was found that the resonance curves demonstrate features of Duffing oscillator, reflecting the nonlinear nature of Landau-Lifshitz-Gilbert-Josephson (LLGJ) system of equations.  There is a critical damping value at which anomalous dependence  comes into play. This critical value depends on the system parameters. Here we present the details of such transformation from usual to anomalous dependence with variation in spin-orbit coupling and ratio of Josephson to magnetic energy.

\begin{figure}
	\includegraphics[height=70mm]{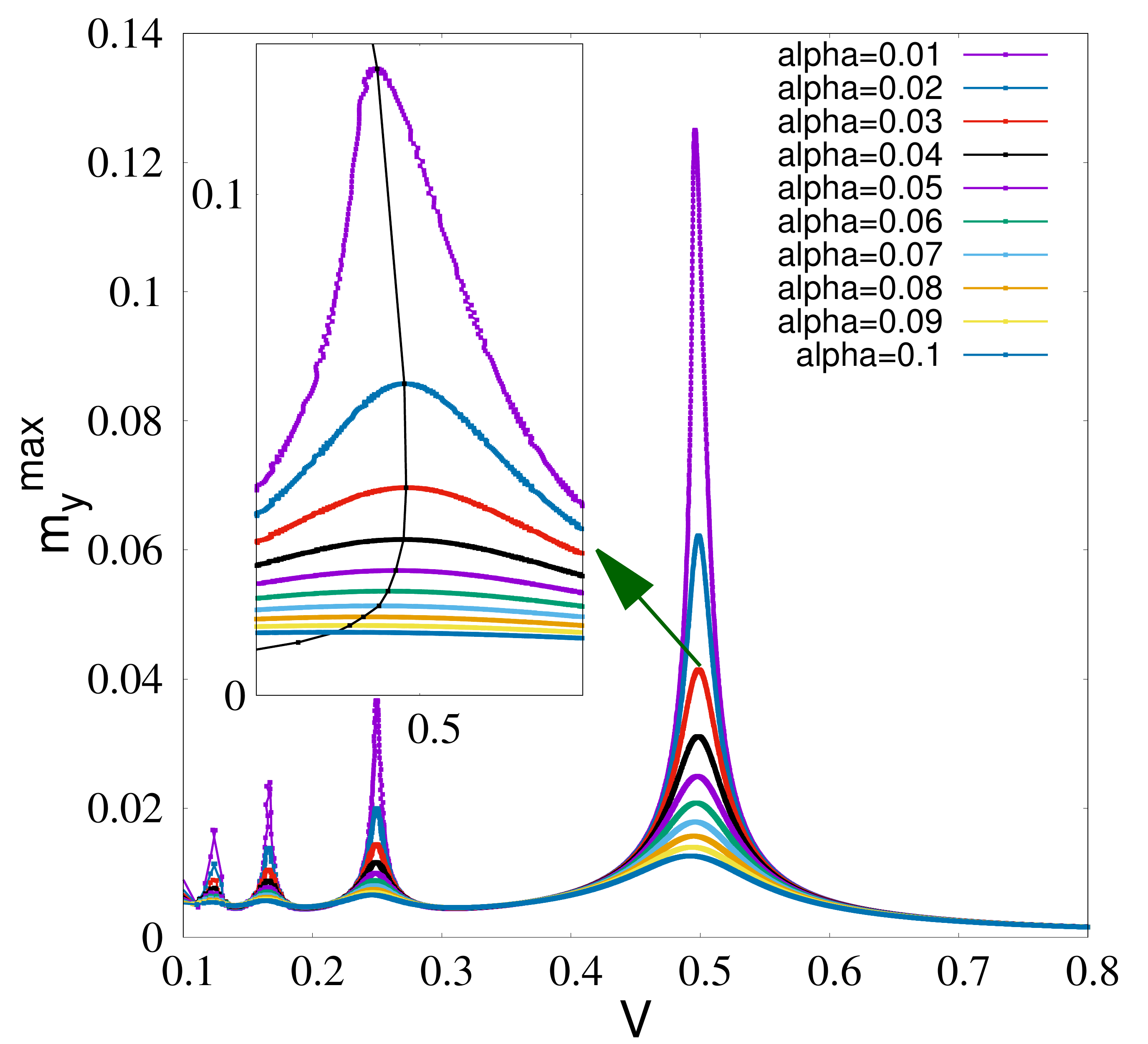}
	\caption{Maximal amplitude of magnetization $m_y-$component at each values of bias current and voltage along IV-characteristics of the $\varphi_0$   junction in the ferromagnetic resonance region for various $\alpha$. Inset enlarges the main maximum. Parameters:  $\beta_c=25$,G=0.05,r=0.05,$\omega_F=0.5$.}
	\label{2}
\end{figure}

To investigate the effect of damping, we calculate the maximal amplitude of magnetization component $m_y$ taken at each value of the bias current  based on the LLGJ system of equations (\ref{syseq}). In Fig.\ref{2} we show the voltage dependence of maximal amplitude $m_y^{max}$  in the ferromagnetic resonance region at different damping parameter and small values of Josephson to magnetic energy ratio G=0.05 and spin-orbit coupling $r=0.05$. We found that the ferromagnetic resonance curves demonstrate the different forms. An increase in damping shows a nonuniform change in the resonant frequency: it is approaching the $\omega_F$ instead of moving away with increase in $\alpha$. We stress that this happens at small $G$ and $r$. We consider that such behavior can be explained by the nonlinear nature of the LLGJ system of equations. There is a manifestation of subharmonics of the FMR in Fig.\ref{2} at $\omega=1/2,1/3, 1/4$.

We usually expect the resonance peak to move away from resonance as the $\alpha$ increases. Figure \ref{2} shows that this normal effect is accompanied with an anomalous behaviour as can be seen in the inset to this figure, where the resonance peak approaches $\omega_F$ as $\alpha$ increases \cite{shukrinov-prb21}.

\begin{figure}
	\includegraphics[height=70mm]{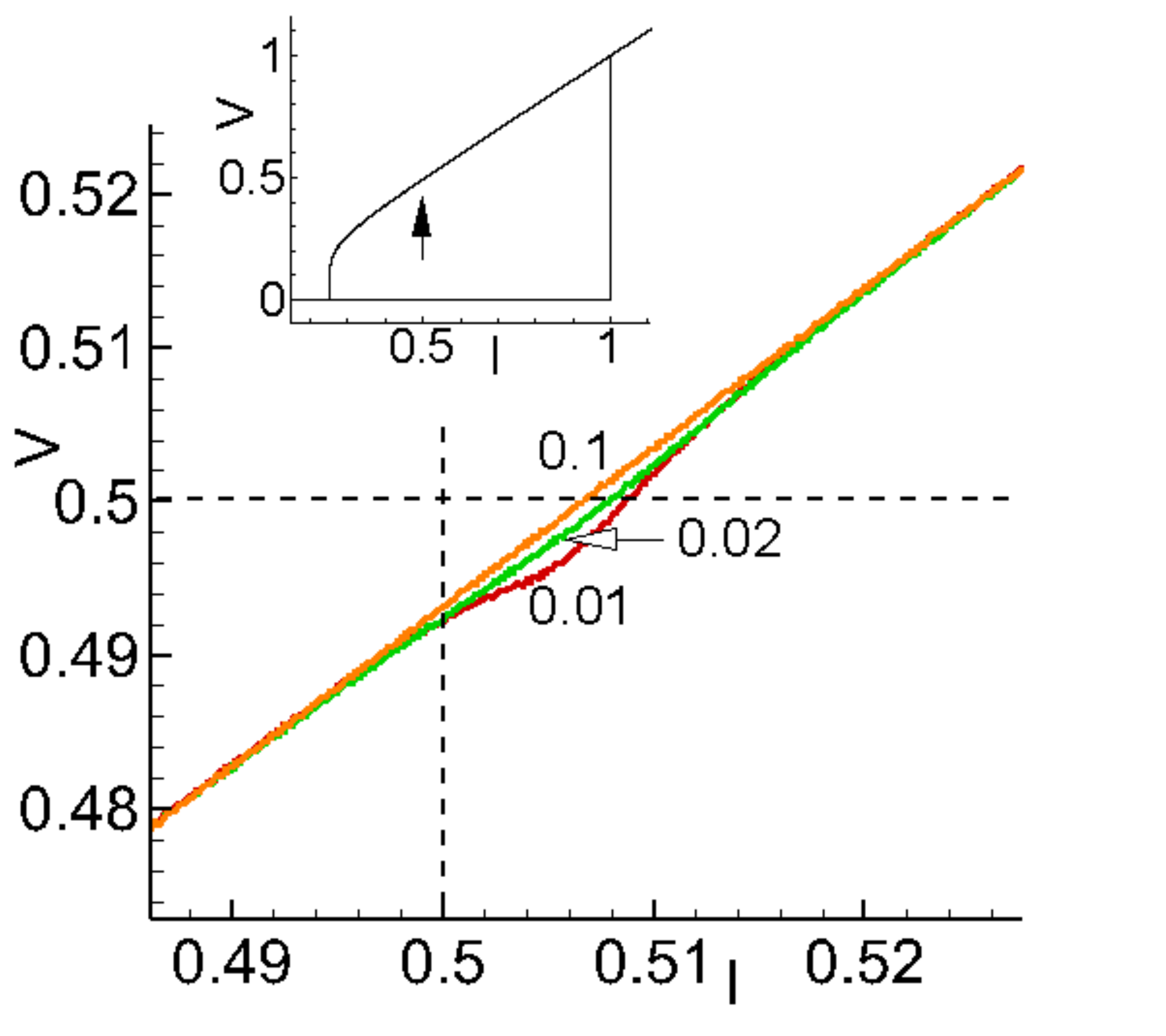}
	\caption{Part of the IV characteristic of the $\varphi_0$ junction at $G=0.05,r=0.05$ and different values of Gilbert damping. The numbers show $\alpha$ value. Inset shows the total IV-characteristic and arrow indicates the resonance region}
	\label{3}
\end{figure}

The manifestation of FMR in IV-characteristics of  the $\varphi_0$ junction at three values of damping parameter is demonstrated in Fig. \ref{3}.  The  strong deviation of the IV-curve is observing at $\alpha=0.01$, which is characteristic value for many magnetic materials. This fact indicates that ADD can be observed experimentally by measuring IV-characteristics in wide interval of the damping  parameter.

Interesting features of ADD appear by a variation of  spin-orbit coupling. As it was demonstrated in Ref.\cite{abdelmoneim22cond-mat}, an increase in SOC leads to the essential change in IV-characteristics and magnetization precession in the ferromagnetic resonance region. The nonlinearity is going stronger and the state with negative differential resistance appears at large SOC.

\begin{figure}
	\includegraphics[height=70mm]{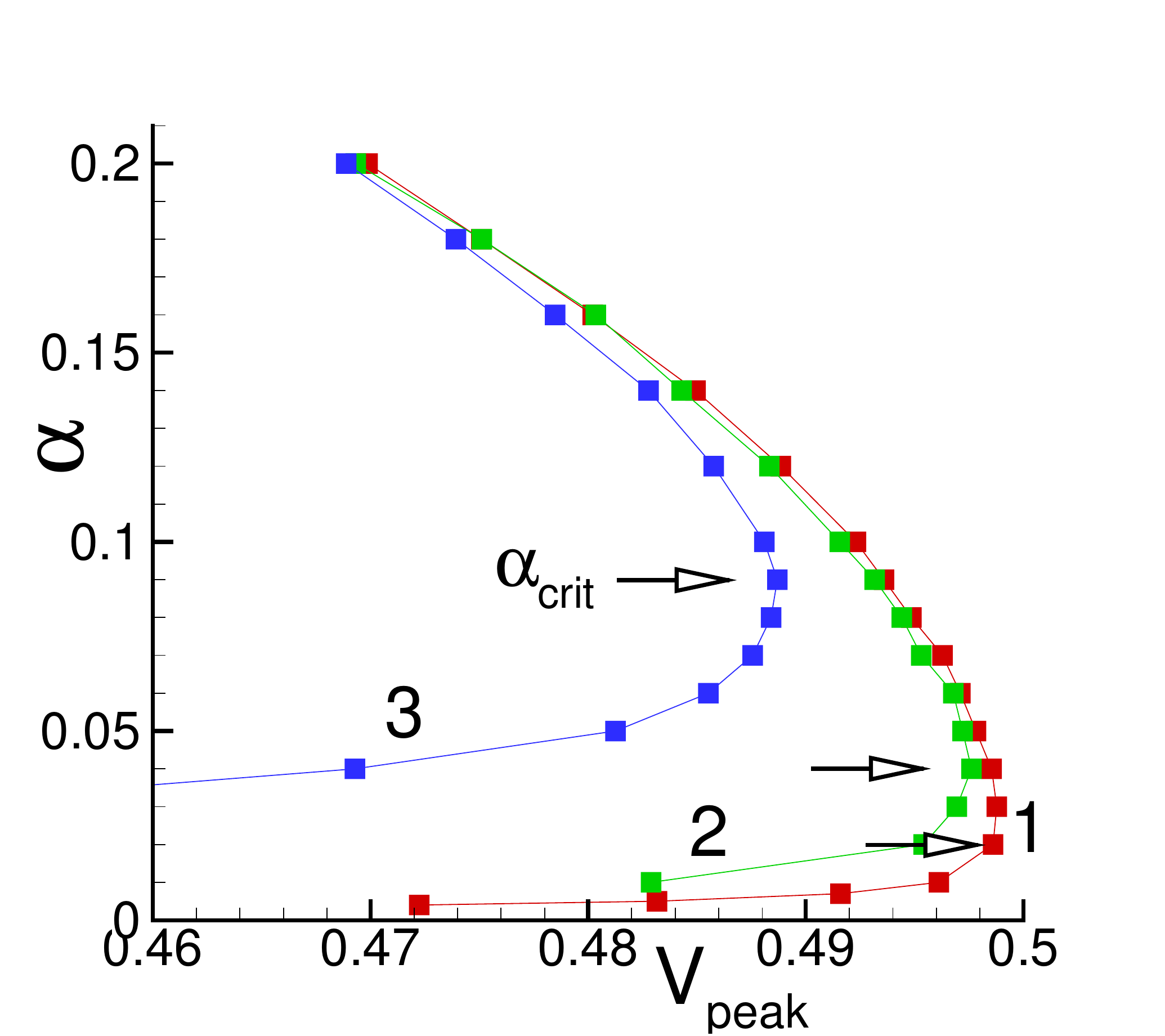}
	\includegraphics[height=70mm]{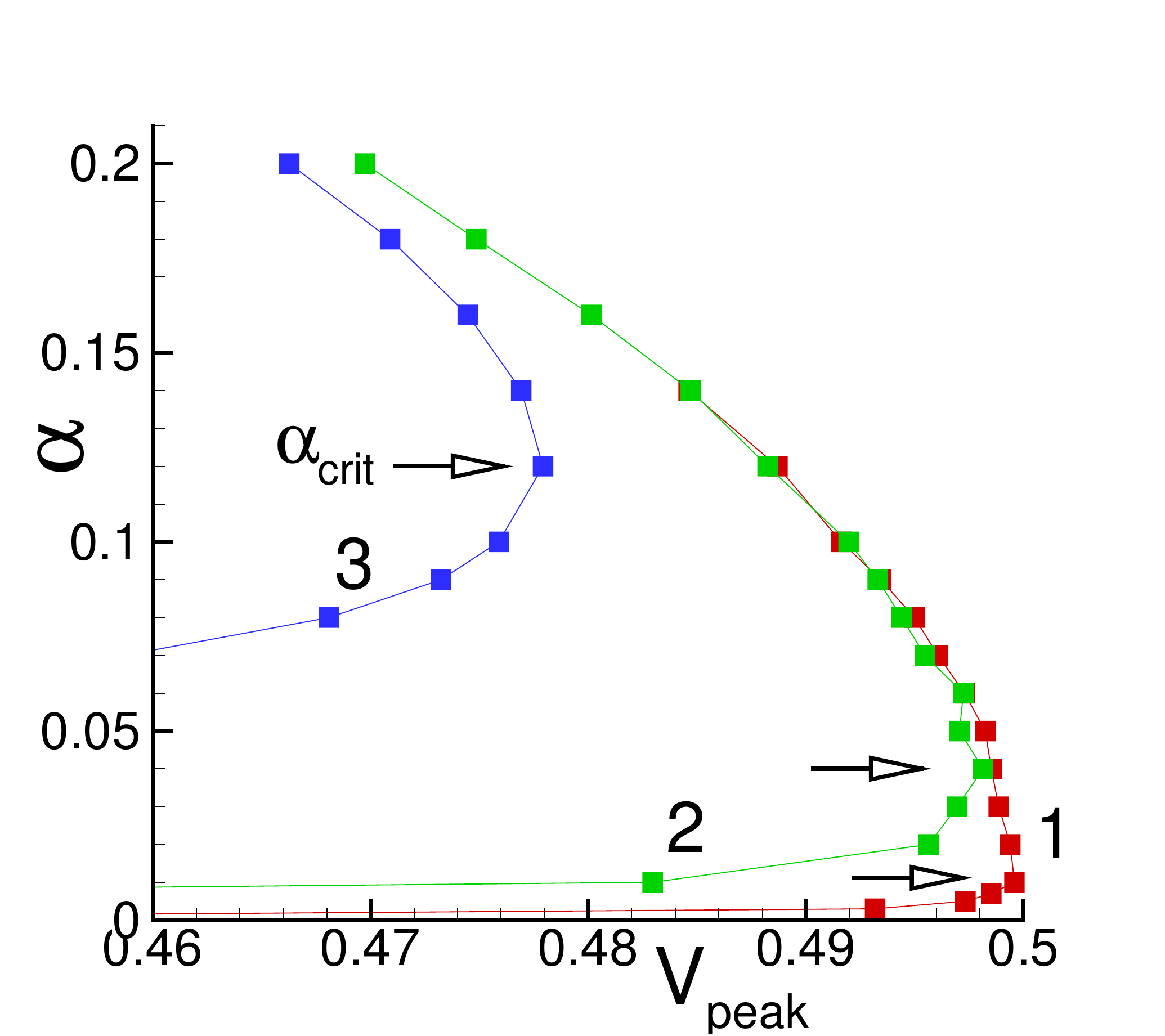}
	\caption{(a) Demonstration of ADD at different values of SOC parameter $r$ at $G=0.05$. Numbers indicate: 1 - $r=0.05$; 2 -  $r=0.1$; 3 - $r=0.5$; Arrows show critical $\alpha$ value, corresponded to the reversal in the $\alpha$ dependence (b) Demonstration of ADD at different values of the Josephson to magnetic energy ratio $G$ at $r=0.05$.  Numbers indicate: 1 - $G=0.01$; 2 -  $G=0.1$; 3 - $G=1$.}	
	\label{4}
\end{figure}

Figure \ref{4}(a) demonstrates results of numerical simulations of $m_y^{max}$ dependence  on $\alpha$ at different values of SOC parameter $r$.
It shows two specific features of ADD.  First,  with an increase in $r$, the critical value of $V_{peak}$ is decreasing (the curve moves away from $\omega_F$). The second important feature is an increasing of $\alpha_{crit}$ which is demonstrated in this figure by arrows.

Another model parameter which affects the phenomenon discussed in the present paper is the ratio $G$ of Josephson to magnetic energies. Figure \ref{4}(b) demonstrates the results of numerical simulations of $m_y^{max}$ dependence  on $\alpha$ at different values of G.

Similar to the effect of $r$, increasing $G$ also causes the value of $\alpha_{crit}$ to increase. By changing the volume of the ferromagnetic layer, the ferromagnetic energy and consequently the value of G can be changed \cite{konschelle-prl09}. For small G, i.e. a situation where the magnetic energy is much larger than the Josephson energy, the magnetic layer receives less energy, and its amplitude decreases in the y direction, and also the maximum value of the oscillation frequency is closer to the magnetic frequency, $\omega_F$.

\subsection{B. Dynamics and IV-characteristics of the $\varphi_{0}$ junction at small system parameters }
As it was discussed in Refs.\cite{konschelle-prl09,shukrinov-pepan20,shukrinov-ltp20}, in case of $G,r,\alpha<<1$ and $m_z\approx1$, first three equations of the system (\ref{syseq}) can be simplified. Taking into account $\varphi=\omega_{J}t$ and neglecting quadratic terms of $m_x$ and $ m_y $,  we get
\begin{equation}
	\label{eq_sys}
	\left\{\begin{array}{ll}
		\displaystyle \dot{m}_{x}=\omega_{F}[-m_{y}+Gr\sin(\omega_{J} t)-\alpha m_{x}]
		\vspace{0.2 cm}\\
		\displaystyle \dot{m}_{y}=\omega_{F}[m_{x}-\alpha m_{y}],
	\end{array}\right.
\end{equation}
This system of equations can be written as the second order differential equation with respect to $m_{y}$
\begin{equation}
	\label{eq_d2my_linear}
	\displaystyle \ddot{m}_{y}+ 2\alpha\omega_{F}\dot{m}_{y}
	+\omega_{F}^{2}m_{y}=
	\omega_{F}^{2}Gr\sin\omega_{J} t.
\end{equation}
Corresponding solution for $m_{y}$ has the form
\begin{equation}
	\label{solution2}
	m_{y}(t)=\frac{\omega_{+}-\omega_{-}}{r}\sin(\omega_{J} t)-\frac{\gamma_{+}+\gamma_{-}}{r}\cos(\omega_{J} t),
\end{equation}
where
\begin{equation}
	\label{coef1}
	\omega_{\pm}=\frac{Gr^{2}\omega_{F}}{2}\frac{\omega_{J}\pm\omega_{F}}{\Omega_{\pm}},
\end{equation}
and
\begin{equation}
	\label{coef2}
	\gamma_{\pm}=\frac{Gr^{2}\omega_{F}}{2}\frac{\alpha\omega_{J}}{\Omega_{\pm}}.
\end{equation}
with $\Omega_{\pm}=(\omega_{J}\pm\omega_{F})^{2}+(\alpha\omega_{J})^{2}$ (see Ref.\cite{konschelle-prl09} and corresponded Erratum\cite{konschelle-prl19}).

When the Josephson frequency $ \omega_{J}$ is approaching  the ferromagnetic one $ \omega_{F} $, $m_{y}$ demonstrates the damped ferromagnetic resonance. Differential resistance in the resonance region is decreasing and it is manifested  in the IV--characteristic as a resonance branch \cite{shukrinov-prb19}.

Taking into account $r m_{y}<<1$, we rewrite expression for superconducting current as
\begin{eqnarray}
	\label{I0current1}
	\displaystyle I_{s}(t)&=&\sin(\omega_{J}t-rm_{y}(t)) \nonumber\\
	&=&\sin(\omega_{J} t)-rm_{y}\cos(\omega_{J}t)
\end{eqnarray}

Using solution (\ref{solution2}) we can obtain
\begin{eqnarray}
	\label{I0current3}
	\displaystyle I_{s}(t)&=&\sin\omega_{J}t-\frac{\omega_{+}-\omega_{-}}{2}\sin2\omega_{J} t\nonumber\\
	&+& \frac{\gamma_{+}+\gamma_{-}}{2}\cos2\omega_{J} t
	+ I_{0}(\alpha)
\end{eqnarray}
where

\begin{eqnarray}
	\label{I0_expression}
	I_{0}=\frac{\gamma_{+}+\gamma_{-}}{2}.
\end{eqnarray}

This superconducting current explains the appearance of the resonance branch in the IV--characteristic. The generated current $I_{0}$ can be expressed through the amplitude of $m_y$ and SOI parameter $r$
\begin{eqnarray}
	\label{I0_expression1}
	I_{0}=\frac{r}{2}m_{y}^{max}(\omega_{J}),
\end{eqnarray}
with $m_{y}^{max}(\omega_{J})$ being the frequency response of $m_{y}$.

\begin{figure}
	\includegraphics[height=50mm]
	{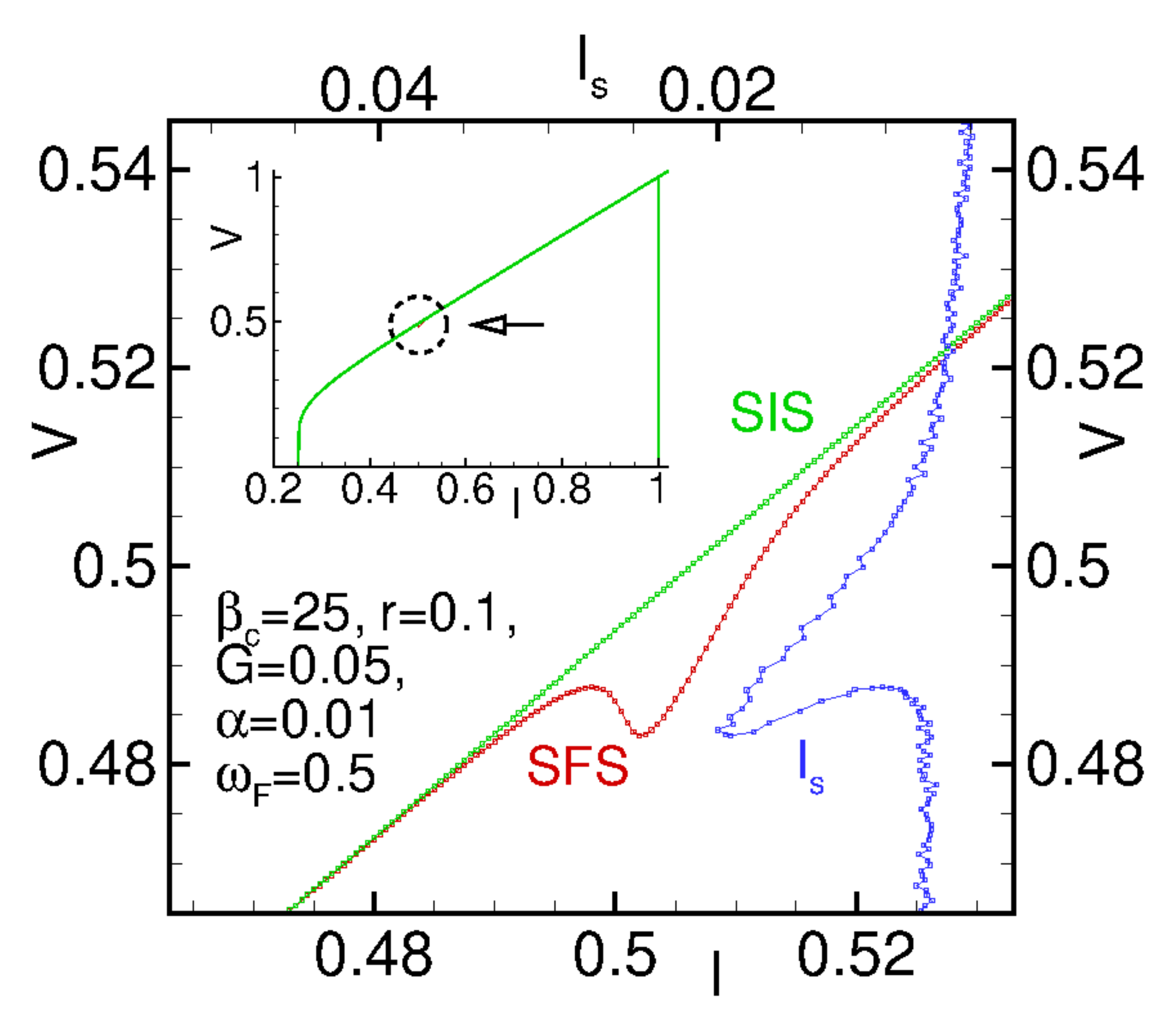}
	\caption{IV-characteristic of $\varphi_{0}$ and SIS junctions and calculated average superconducting current through the $\varphi_{0}$ junction.}
	\label{5}
\end{figure}

At small model parameters $\alpha<<Gr<<1$ of SFS $\varphi_0$ Josephson junction, the states with a negative differential resistance  appear in  the IV-characteristics in the FMR region. Due to the nonlinearity, the resonance peak is asymmetric. Increasing of nonlinearity leads to the bistability (foldover effect). A natural questions appears if the states with a negative differential resistance are the origin of the foldover and ADD. In order to clarify this question, we show in Fig.\ref{5} a part of IV--characteristics of $\varphi_{0}$ junction  together with IV-characteristic of SIS junction in the ferromagnetic resonance region and numerically calculated superconducting current through this junction. The total IV--characteristics are demonstrated in the inset to this figure.

We see the correlation of the foldover effect in superconducting current (blue) with the NDR part of IV-curve. The peak in the superconducting current and minimum of IV-curve are at the same voltage value. So, both effects reflect the nonlinear features of the ferromagnetic resonance in $\varphi_{0}$ junction. But in contrast to the foldover and ADD effects, which start manifesting themselves at relatively small deviations from linear case,  the nonlinearity in case of the NDR plays a more essential role.

We note that in the resonance region for considered limit of model parameters  the $m_{y}$ amplitude are coupled to the value of superconducting current (see Eq.(\ref{I0_expression1})). We stress the importance of the  performed analysis demonstrating  the analytical coupling  of time independent superconducting current and magnetization, reflecting the  Duffing oscillator features of the $\varphi_{0}$ junction.

As it is well known, the states with negative differential resistance appear in IV-characteristics of Josephson structures in different physical situations. In particular, the nonlinear superconducting structures being driven far from equilibrium exhibit NDR states  \cite{Pedersen_2014}. The NDR states plays an essential role in the applications related to the THz radiation emission \cite{Kadowaki_2008}. A detailed explanation for different types of negative differential resistance (NDR) in Josephson junction (i.e., N-shaped and S-shaped) is introduced in\cite{filatrella2014}. The authors  emphasize that the nonlinear behavior of the Josephson junction plays a key role in the NDR feature. In our case, the NDR states appear as a result of system's nonlinearity at small values of $\varphi_0$ junction parameters, such as SOC, ratio of Josephson to magnetic energy and Gilbert damping.  We demonstrate these effects here by by presented results of detailed investigation of the NDR state at different system parameters and discuss possibility of their control near the ferromagnetic resonance.

\begin{figure}
	\centering
	\includegraphics[width=7cm]{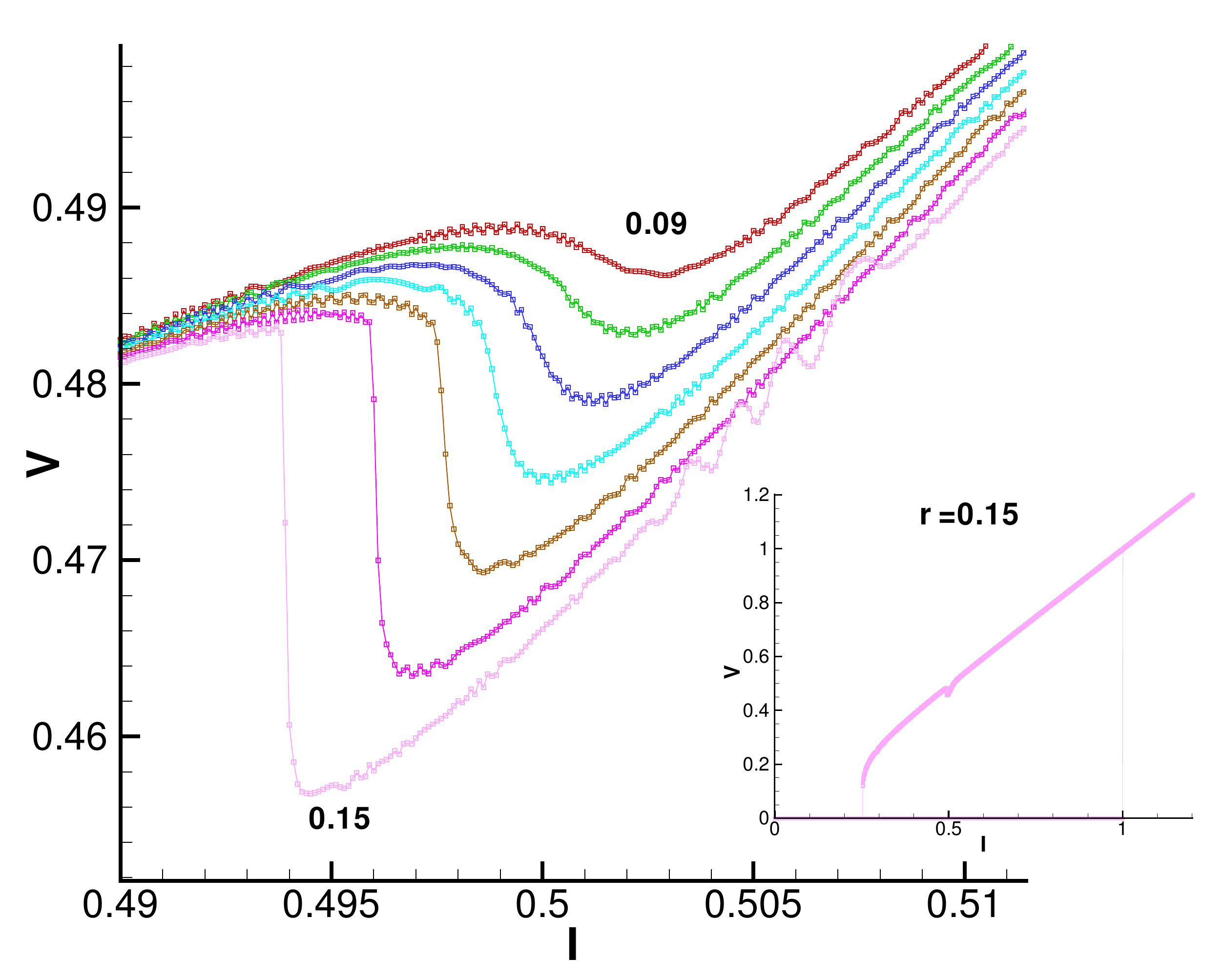}
	\caption{Enlarged parts of the IV-curves  in the resonance region at different values of SOC parameter $r$  at $\alpha=0.01$ and $G=0.05$. The numbers indicate the increasing of $r$ from 0.09 to 0.15 with an increment 0.01. Inset demonstrates total IV-characteristic ar $r=0.15$} \label{6}
\end{figure}

Figure \ref{6} shows the effect of the spin-orbit coupling on the IV-characteristic at $G=0.05$ and $\alpha=0.01$. We see the  NDR feature which is getting more pronounced with increase in $r$.  A further increase in $r$ leads to the jump down in voltage and then  practically linear growth of IV-characteristic.

\begin{figure}
	\centering
	\includegraphics[width=6cm]{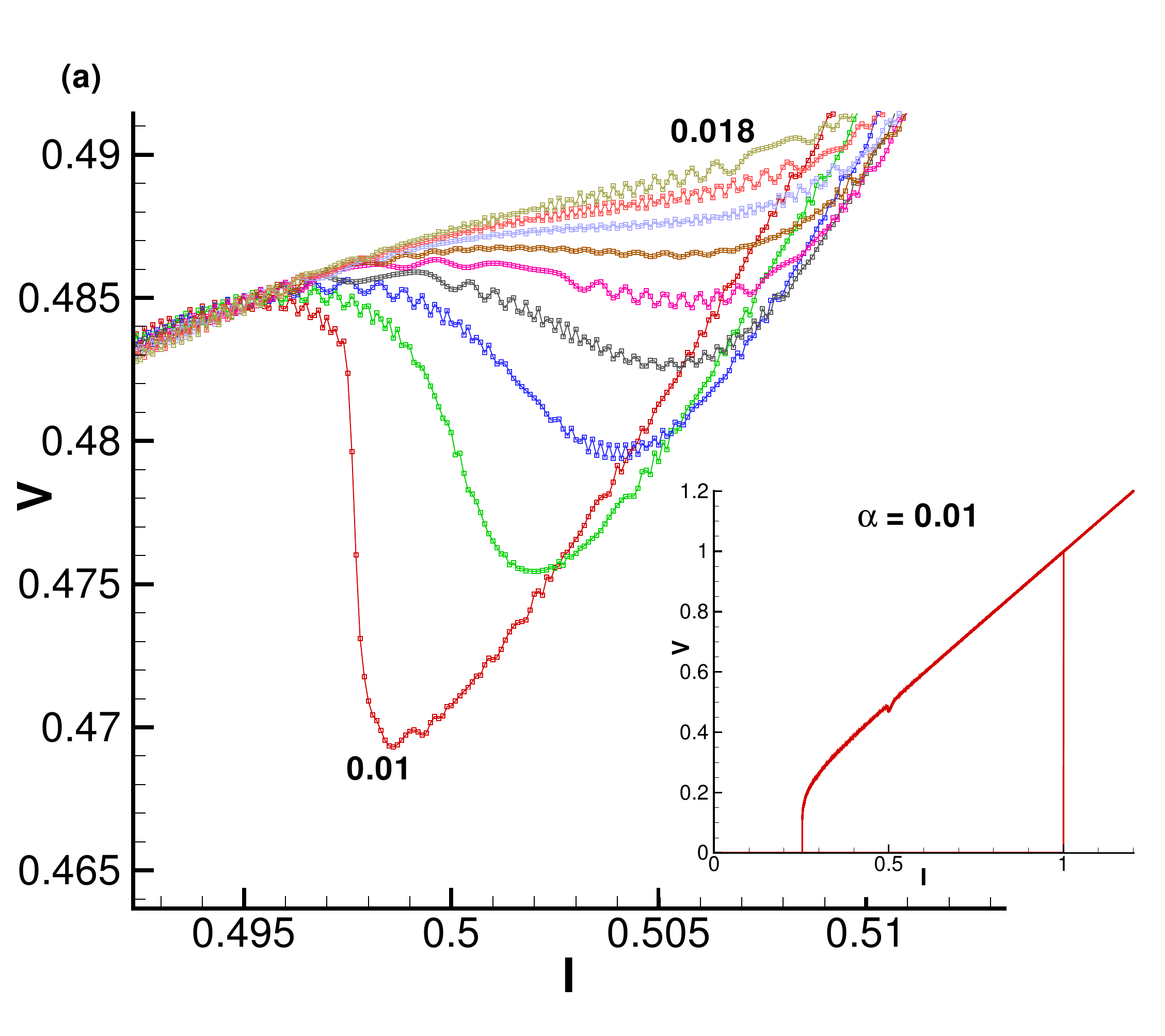}
	\includegraphics[width=6cm]{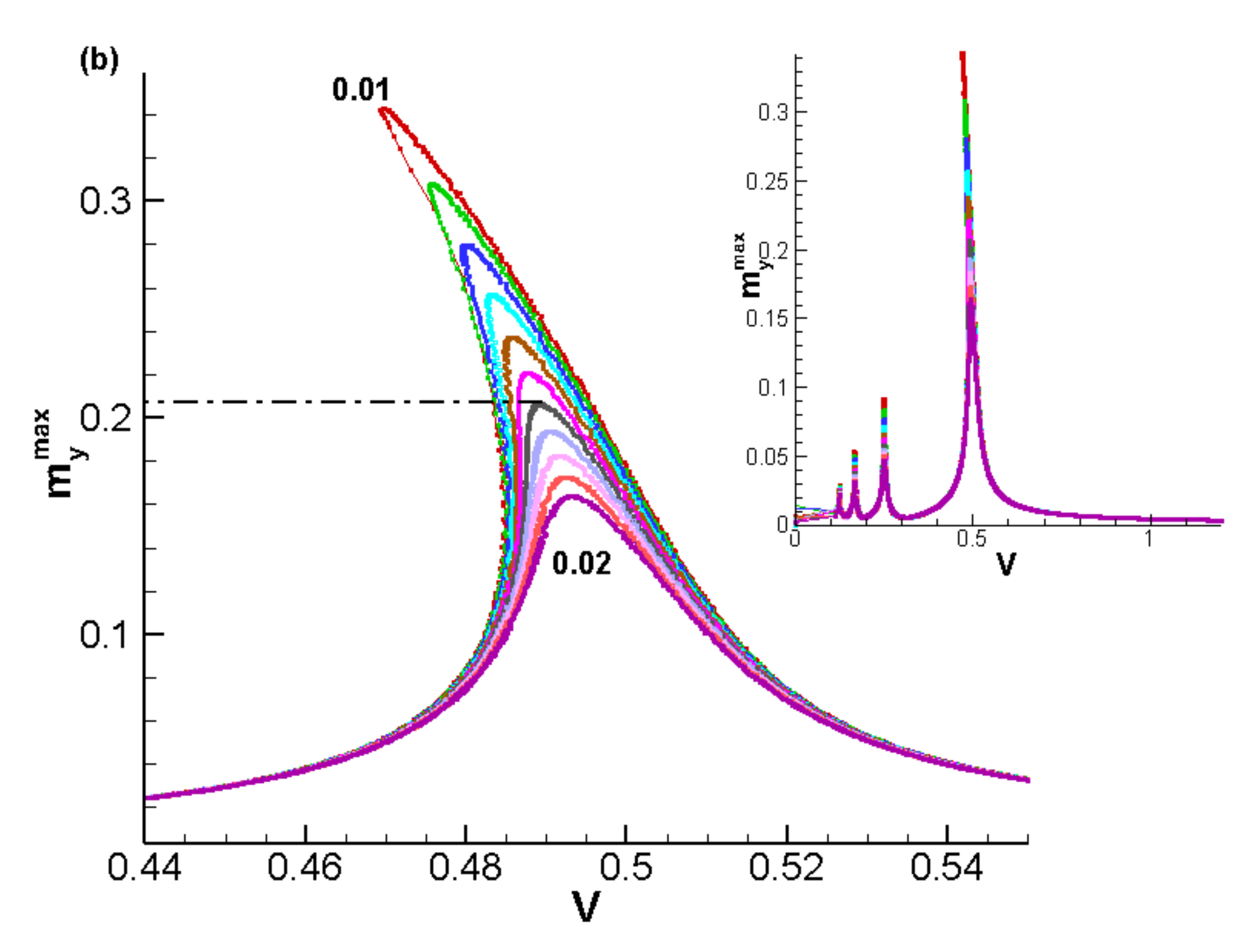}
	\includegraphics[width=6cm]{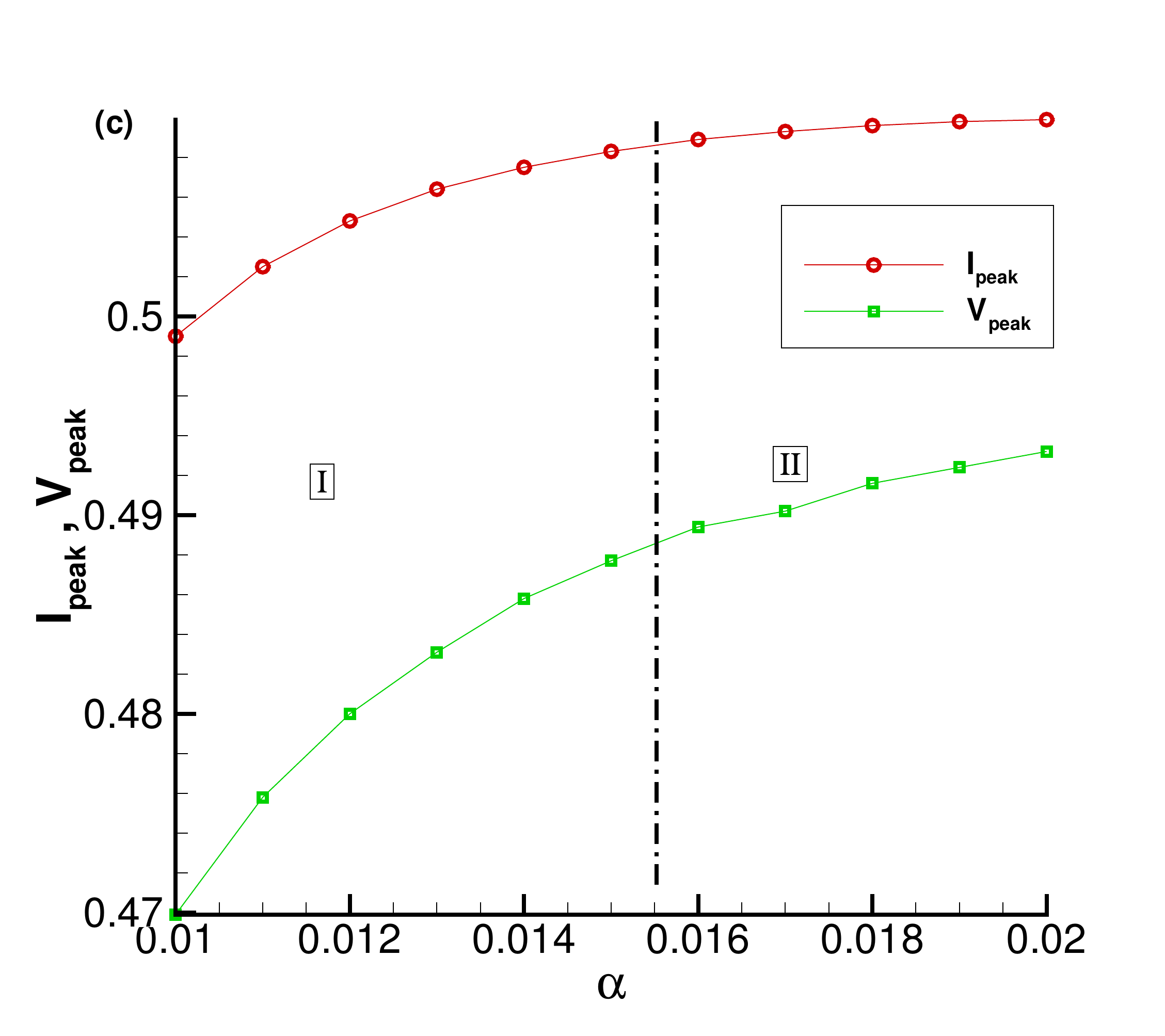}
	\caption{(a) Enlarged parts of the IV-curves at different values of $\alpha$; (b) Voltage dependence of $m^{max}_{y}$ at different $\alpha$; (c) $\alpha$-dependence of the resonance curve maximum  in current ($I_{peak}$) and voltage ($V_{peak}$). The numbers indicate the value of $\alpha$ from 0.01 to 0.018 in (a) and from 0.01 to 0.02 in (b) by an increment 0.001. Results are obtained at $r=0.13$, and $G=0.05$.} \label{7}
\end{figure}

An interesting question is concerning the effect of Gilbert damping. Results of IV-characteristics simulations  in the resonance region at a certain range of the damping parameter $\alpha$ at $G=0.05$ and $r=0.13$ are shown in Fig.\ref{7}(a). In this case, the most pronounced characteristic appears at $\alpha=0.01$. At chosen set of parameters $G=0.05$ and $r=0.13$,  the range of $\alpha$ with pronounced NDR features  is $0.01 \leqslant \alpha < 0.014$.

The maximal amplitude $m^{max}_{y}$ as a function of voltage is shown  in the Fig.\ref{7}(b). Based on the results, presented in Fig.\ref{7}(a) and Fig.\ref{7}(b), we came to the important conclusion  that the foldover effect (bistability) and NDR state have strong correlations and have the same origin related to  the nonlinearity at small system parameters.

But anomalous damping dependence does not show a one-to-one correlation with either negative differential resistance or foldover effect. The resonance peak positions of $m_y^{max}$  in bias current $I_{peak}$, and in voltage $V_{peak}$  as the functions of $\alpha$ are demonstrated in Fig.\ref{7}(c). According to our results, we can divide $\alpha$ interval into two regions (see Fig.\ref{7}(c)). Region I introduces the values of  $\alpha$ where the NDR feature is present, while in the Region II it disappears. In the Region II the  foldover effect (bistability) disappears as well, but ADD is realized.

\subsection{C. Duffing oscillator features of $\varphi_{0}$ junction and critical damping}

The system of equations (\ref{syseq}) is nonlinear and very complex, so in order to provide analytical study of dynamics of the
$\varphi_{0}$ junction we need to derive an approximated equation for some limited values of model parameters. In Ref.~\cite{shukrinov-prb21} was shown that the resonance curves demonstrate features of Duffing oscillator, reflecting the nonlinear nature of Landau-Lifshitz-Gilbert-Josephson (LLGJ) system of equations. In this section, we present analytical approach to describe the nonlinear dynamics of $\varphi_{0}$-junction and compare analytical results followed from approximated Duffing equation with numerical simulations of total system of equations (\ref{syseq}). We show that in the limit of $\alpha<<G,r<<1$, we arrive to the Duffing oscillator.
We start by first three equations of the (\ref{syseq}) for magnetization components
\begin{eqnarray}
	\label{eq_LLG1}
	\frac{\dot{m}_{x}}{\omega_{F}}&=&-m_{y}m_{z}+Grm_{z}\sin(\varphi-rm_{y})-\alpha m_{x}m_{z}^{2}\nonumber\\
	\frac{\dot{m}_{y}}{\omega_{F}}&=&m_{x}m_{z}-\alpha m_{y}m_{z}^{2}\\
	\frac{\dot{m}_{z}}{\omega_{F}}&=&-Grm_{x}\sin(\varphi-rm_{y})+\alpha m_{z}(m_{x}^{2}+m_{y}^{2})\nonumber
\end{eqnarray}

Simplifying this system of equations by the same procedure as it was done in Ref.\cite{shukrinov-prb21} we can write equation for $m_{y}$ as

\begin{equation}
	\label{eq_d2my_li_alpha4}
	\displaystyle
	{\ddot m_y} + 2 \xi \alpha \dot m_y + \xi^2 (1+\alpha^2)m_y - \xi^2 (1+\alpha^2 - \alpha^4)m_y^3 = \xi^2 \gamma \sin \varphi.
\end{equation}

Finally, by neglecting the $\alpha^2$ and $\alpha^4$ terms which are much smaller than  1, we come to well known Duffing equation,

\begin{equation}
	\label{eq_d2my_li}
	\displaystyle
	{\ddot m_y} + 2 \omega_{F} \alpha \dot m_y + \omega_{F}^2 m_y - \omega_{F}^2 m_y^3 = \omega_{F}^2 Gr \sin \varphi.
\end{equation}

In the small parameter range, this Duffing equation can describe the dynamics of $m_y$. We will have the full dynamics, once we consider the coupling with the Josephson equation,

\begin{equation}
	\label{eq_josephson_single}
	\displaystyle
	{\ddot \varphi} + {1 \over {\beta_c}} [{\dot \varphi}-r{\dot m}_{y} + \sin(\varphi - rm_y)] = {1 \over {\beta_c}} I.
\end{equation}
The system of equations, (\ref{eq_d2my_li}) and (\ref{eq_josephson_single}) can replace the LLGJ equations in the limit of $G,r<<1$ and $G,r>>\alpha$.

Taking into account $\varphi=\omega_{J}t$ we can right analytically obtained frequency response for equation (\ref{eq_d2my_li})

\begin{eqnarray}
	\label{w_response}
	(m_{y}^{max})^{2}=\frac{\big(Gr\big)^{2}}{\big[\omega^{2}-1+\frac{3}{4}(m_{y}^{max})^{2}\big]^{2}+\big(2\alpha\omega\big)^{2}}\nonumber\\
\end{eqnarray}
where $\omega=\omega_{J}/\omega_{F}$. From Eq. (\ref{w_response}) we get

\begin{eqnarray}
	\label{eq_cubic}
	(m_{y}^{max})^{6}&+&\frac{8}{3}(\omega^{2}-1)(m_{y}^{max})^{4}\nonumber\\
	&+&\bigg(\frac{4}{3}\bigg)^{2}\bigg[(\omega^{2}-1)^{2}+\big(2\alpha\omega\big)^{2}\bigg](m_{y}^{max})^{2}\nonumber\\
	&-&\bigg(\frac{4}{3}Gr\bigg)^{2}=0.
\end{eqnarray}

\begin{figure}
	\includegraphics[height=50mm]{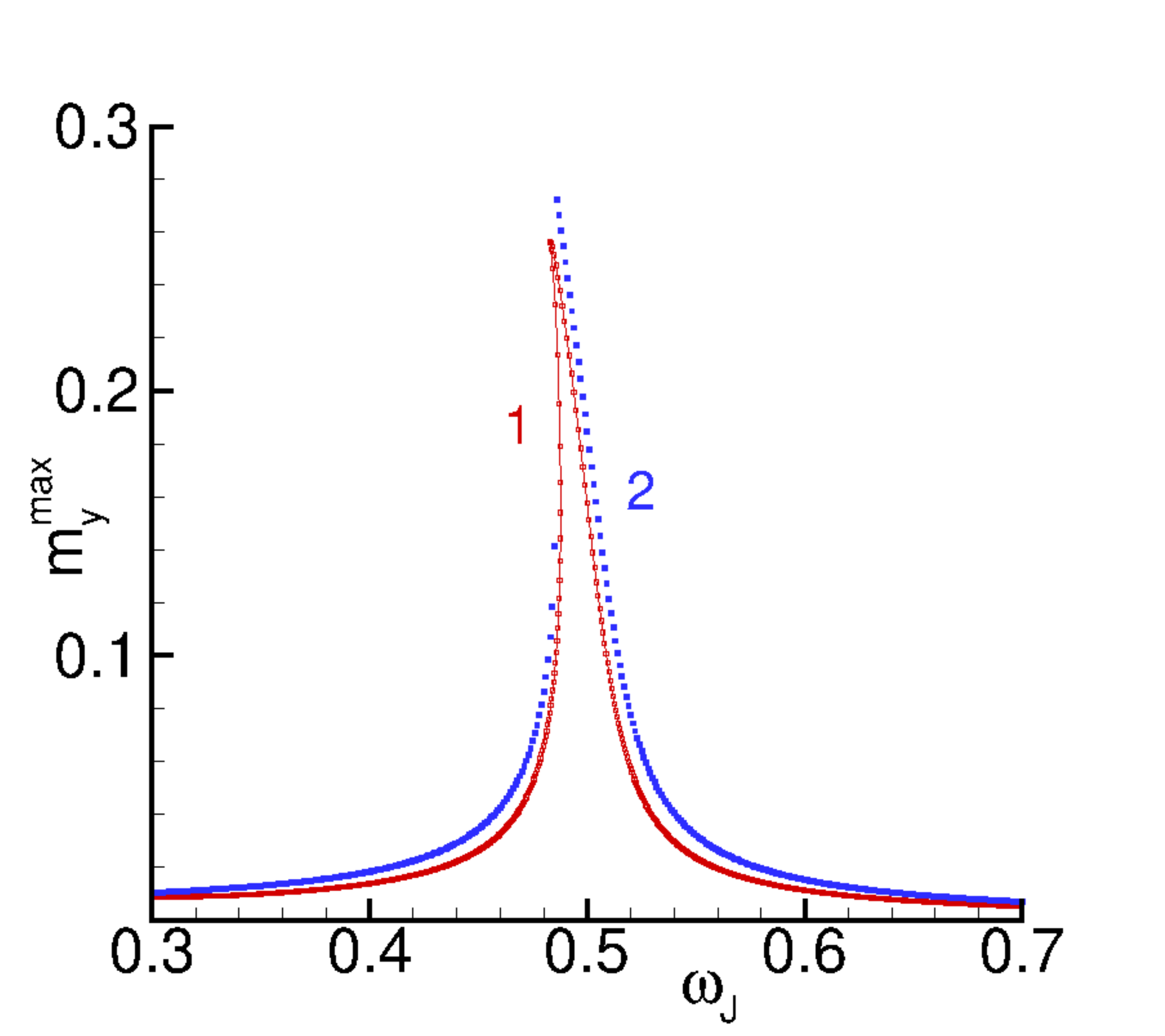}
	\caption{Numerically (curve 1) and analytically (curve 2) calculated amplitude dependence of $m_{y}$.}
	\label{8}
\end{figure}

This equation allows to determine analytically  frequency dependence of the $m^{max}_{y}$ amplitude. To find it we solve the equation (\ref{eq_cubic}) by the Newton method. Results of analytical calculations (blue dots) corresponded  to (\ref{eq_cubic}) and numerical one (red doted line) corresponded to the full system of equation (\ref{syseq}) are demonstrated in Fig.\ref{8}.

We can see that they are close to each other which proves the correctness of the chosen approximation. Both curves demonstrate an asymmetric resonance peak, which is common for Duffing oscillator. When a role of the cubic term is getting larger, we observe a bistability  of the resonance curve, which is usually called a foldover effect. Note that the foldover effect can be also achieved by the damping decreasing; i.e., by the decreasing of dissipative term in (\ref{eq_d2my_li}), we can increase the influence of the cubic term in this equation.

\begin{figure}
	\includegraphics[height=50mm]{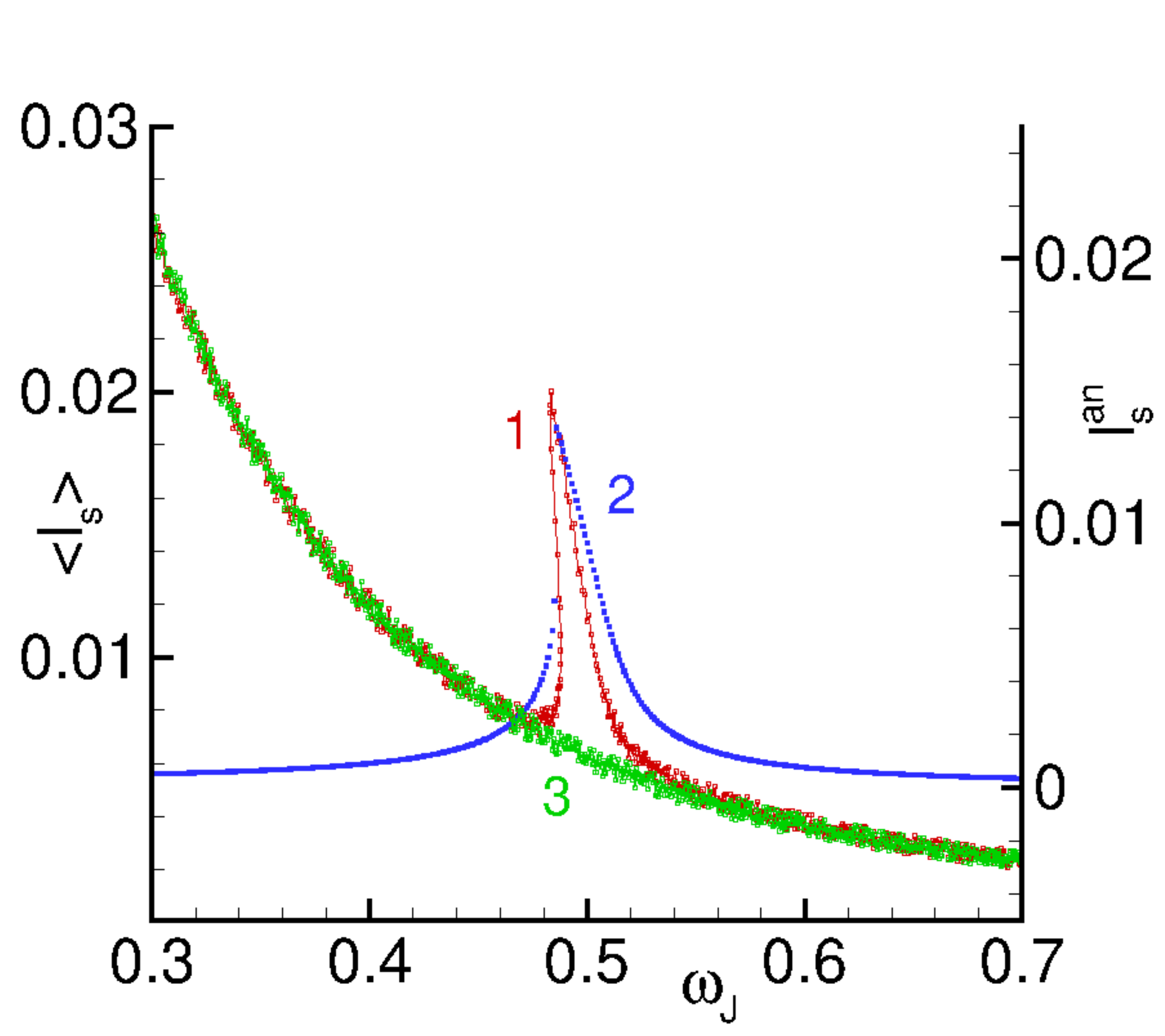}
	\caption{Numerically calculated superconducting current for SFS junction (plot 1) and analytical $I_{0}$ (plot 2) and superconducting current for SIS junction (plot 3).}
	\label{9}
\end{figure}

The comparison of analytically and numerically calculated superconducting current as a function of the Josephson frequency is demonstrated in Fig. \ref{9}. We note that in our normalization $V=\omega_J$.
We can see the manifestation of the asymmetric resonance peak in the frequency dependence of superconducting current. So, the approximated system of equations {\ref{eq_sys}} reflects one of the main feature of Duffing oscillator.

\begin{figure}
	\includegraphics[height=70mm]{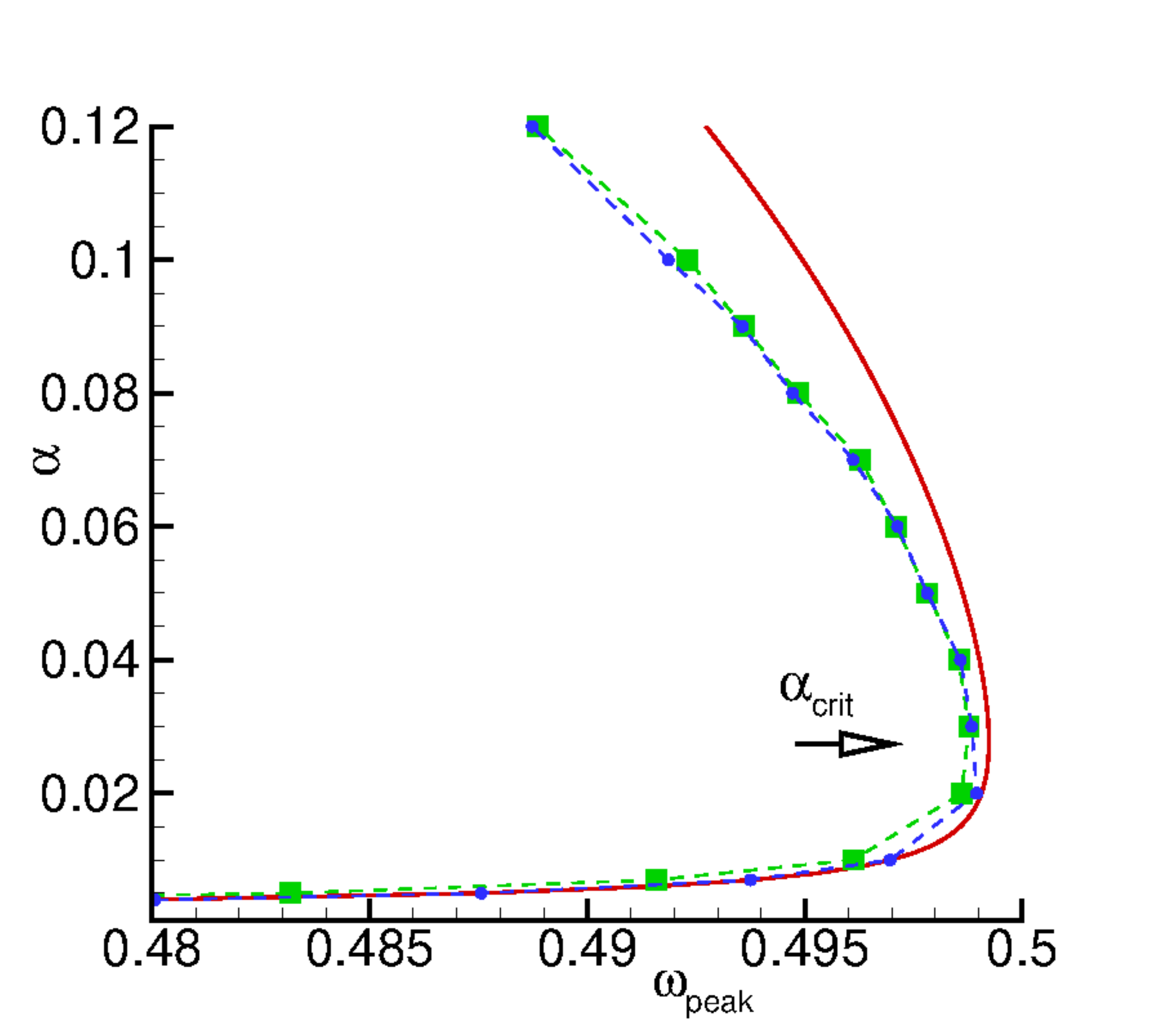}
	\caption{The $\alpha$-dependence of the resonance maximum of $m_{y}^{max}(V)$ in the damping parameter interval [0.001 – 0.12].  Green squares show results calculated numerically  according to the full system of  equations (\ref{syseq}), blue circles show results calculated numerically according to the generalized Duffing and Josephson equations (\ref{eq_d2my_li_alpha4},\ref{eq_josephson_single}). The dashed line connects the symbols to guide eyes. Solid line show analytical $\alpha$-dependence  calculated according to the Eq. (\ref{w_solutionn}). All calculation have been done at $\beta_c=25$, G=0.05, r=0.05, $\omega_F=0.5$. }
	\label{10}
\end{figure}

Figure (\ref{10}) compares anomalous damping dependence of the resonance peak of $m_y^{max}(V)$  calculated numerically  according to the full LLGJ system of  equations (\ref{syseq}) with calculated numerically according to the generalized Duffing model (equations (\ref{eq_d2my_li_alpha4}, \ref{eq_josephson_single})). We see that in the damping parameter interval [0.001 – 0.2] the coincidence of the dependences is enough good.

Using equation (\ref{eq_d2my_li}) with $\varphi=\omega_{J} t$, we can find (see Supplementary materials \ref{si:1}) a relation between position of the resonance peak in $m_y^{max}(V)$ dependence and damping

\begin{eqnarray}
	\label{w_solutionn}
	\omega_{peak}=\sqrt{\frac{1-3\alpha^{2}}{2}+ \frac{1}{2}\sqrt{(1-\alpha^{2})^{2}-12(\frac{Gr}{4\alpha})^{2}}}
\end{eqnarray}
where $\omega_{peak}=\frac{\omega_{J,peak}}{\omega_F}$ determines  the position of the resonance peak.

Equation (\ref{w_solutionn}) allows to find the formula for critical damping $\alpha_{crit}$ which is an important parameter determining the reversal point in damping dependence of the resonance peak of  $m_y^{max}(V)$ .

Taking into account equation (\ref{w_solutionn}) we can write equation with respect of $Gr/(4\alpha)$ (See supplementary materials \ref{si:1}).

\begin{eqnarray}
	\label{alpha_derivaitve1}
	9\bigg(\frac{Gr}{4\alpha_{crit}}\bigg)^{4}&+&3\alpha_{crit}^{2}(10\alpha_{crit}^{2}-1)\bigg(\frac{Gr}{4\alpha_{crit}}\bigg)^{2}\\
	&-&2\alpha_{crit}^{4}(\alpha_{crit}^2-1)^{2}=0\nonumber
\end{eqnarray}

Using approximation $10\alpha_{crit}^{2}<<1$ and $\alpha_{crit}^2<<1$ it gives (see Supplementary Materials)

\begin{eqnarray}
\label{alphacritical_4}
\alpha_{crit}\approx\frac{1}{2}\sqrt{\sqrt{\frac{3}{2}}Gr}
\end{eqnarray}

Table 1: A comparison between the numerical and analytical values of
$\alpha_{crit.}$ at different values of $G$ and $r$.

\begin{center}
	\begin{tabular}{ |c|c|c|c|c| }
		\hline
		G&r  &Gr  &$\alpha_{crit.},numerics$ & $\alpha_{crit.},analytics$\\
		\hline
		\hline
		0.01   & 0.05  & 0.0005  &0.0100     &0.0123\\
		0.05   & 0.05  & 0.0025  &0.0300     &0.0276\\
		0.05   & 0.10  & 0.0050  &0.0400     &0.0391\\
		0.05   & 0.30  & 0.0150  &0.0700     &0.0677\\
		0.05   & 0.50  & 0.0250  &0.0900     &0.0874\\
		0.10   & 0.05  & 0.0050  &0.0391     &0.0391\\
		0.60   & 0.05  & 0.0300  &0.0950     &0.0958\\
		0.70   & 0.05  & 0.0350  &0.1000     &0.1035\\
		1.00   & 0.05  & 0.0500  &0.1200     &0.1237\\
		\hline
	\end{tabular}
\end{center}

\begin{figure}
	\includegraphics[height=70mm]{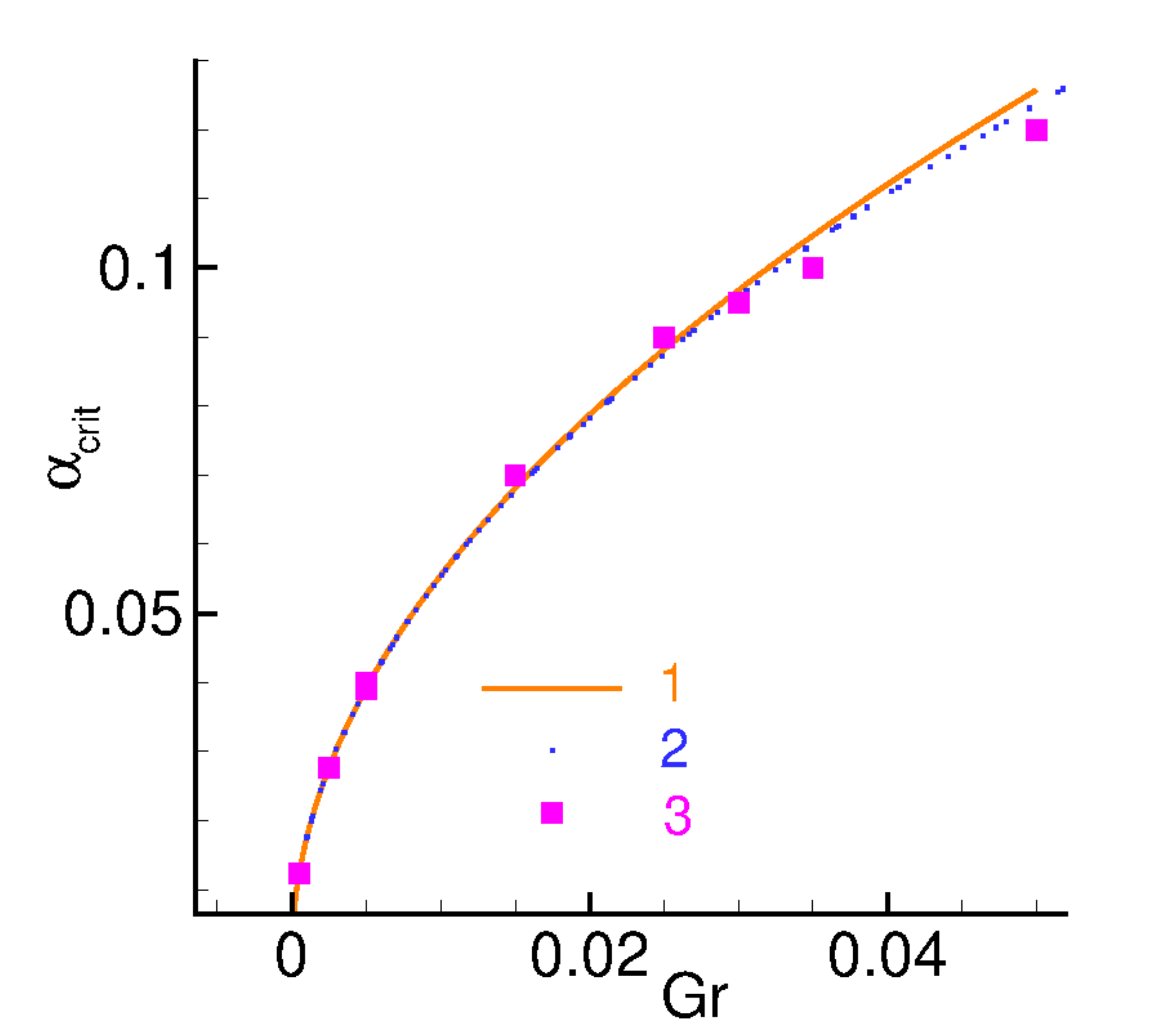}
	\caption{Numerical calculations according to Eq. (\ref{syseq}) (squares), analytical according to Eq.  (\ref{alpha_derivaitve1})(solid line) and approximated analytical according to Eq. (\ref{alphacritical_4}) (dashed line).}
	\label{11}
\end{figure}

Figure \ref{11} presents comparison of numerical and analytical results $\alpha_{crit}$ versus $Gr$.

As we see, it shows a good agreement of numerical and analytical results of calculations at small product of Josephson to magnetic energy ratio  and spin-orbit interaction.

\section{IV. CONCLUSIONS}
The understanding of the nonlinear features of magnetization dynamics in superconductor-ferromagnet-superconductor Josephson junction and their manifestation in the IV-characteristics has implications for superconductor spintronics, and modern information technology. In $\varphi_0$  junctions the nonlinear features can affect the control of magnetization precession by superconducting current and external electromagnetic radiation \cite{{abdelmoneim22cond-mat}}.

Here, using numerical and analytic approaches,  we have demonstrated that at small system parameters, namely, the damping, spin-orbit interaction and Josephson to magnetic energy ratio in $\varphi_0$ junction, magnetic dynamics is reduced to the dynamics of the scalar Duffing oscillator, driven by the Josephson oscillations. We have clarified the role of increasing superconducting current in the resonance region leading to the foldover effect in the ferromagnet magnetization. We have demonstrated the parameter dependence of the anomalous ferromagnetic resonant shifting with anomalous damping dependence due to nonlinearity of the full LLGJ equation  and in its different approximations.  We have derived the analytical expression for critical damping value. Also, we demonstrated appearance of negative differential resistance in the IV-characteristics and the correlation with occurrence of the foldover effect in the magnetization of ferromagnet.

We have stressed that the manifestation of negative differential resistance is related to the nonlinear features of the system\cite{nagel2008, filatrella2014}. It was demonstrated that in the small model parameters case the equation for magnetic subsystem takes form of Duffing equation where nonlinearity manifest itself as the cubic term. We have shown that the appearance of negative differential resistance in the I-V curve is related to the appearance of foldover in the $m_y^{max}$-$V$ curve.


We believe that the experimentally measured IV-characteristics of $\varphi_0$ junction with manifestations discussed in detail in the present paper, would allow close investigations of its nonlinear features important for superconductor electronics and spintronics.

\section{Supplementary}
In supplementary material are presented the details of calculations for Eq.\ref{w_solutionn} and Eq.\ref{alphacritical_4}.

\section{Funding}
Numerical simulations were funded by Project No. 18-71-10095 of the Russian Science Foundation. The presented results concerning the calculations of DC superconducting current in the section V are supported by the Russian Science Foundation in the framework of project 22-42-04408. A.J. and M.R.K. are grateful to IASBS for financial support.

\end{document}